\documentclass[aps,pra,reprint,groupedaddress,showpacs,amsmath,amssymb,twocolumn,floatfix]{revtex4}
\usepackage{txfonts}
\usepackage{graphicx,amssymb,mathrsfs,bm,times,color}
\usepackage{ulem}

\newcommand{\be}{\begin{equation}}

\makeatletter

\newcommand{\Rmnum}[1]{\expandafter\@slowromancap\romannumeral #1@}
\makeatother

\begin{document}

\title {Effective Long-distance Interaction from Short-distance Interaction in Periodically Driven One-dimensional Classical System }

\author{Lingzhen Guo$^{1,2}$}
\author{Modan Liu$^{1,3}$}
\author{Michael Marthaler$^{1}$}
\affiliation{$^1$\mbox{Institut f\"ur Theoretische Festk\"orperphysik, Karlsruhe Institute of Technology (KIT), D-76131 Karlsruhe, Germany}\\
$^2$\mbox{Department of Microtechnology and Nanoscience (MC2),
Chalmers University of Technology, SE-41296 G\"oteborg,
Sweden}\\
$^3$\mbox{Department of Physics, Beijing Normal University,
Beijing 100875, China}\\
}

\date{\today}

\begin{abstract}
We study the classical dynamics of many interacting particles in a
periodically driven one-dimensional (1D) system. We show that
under the  rotating wave approximation (RWA), a short-distance 1D
interaction ($\delta$ function or hard-core interaction), becomes
a long-distance two-dimensional (2D) interaction which only
depends on the distance in the phase space of the rotating frame.
The RWA interaction describes the effect of the interaction on the
slowly changing amplitude and phase of the oscillating particles,
while the fast oscillations take on the role of a force carrier,
which allows for interaction over much larger effective distances.
\end{abstract}


\maketitle

\section{Introduction}

The properties of one-dimensional (1D) interacting particles have been
investigated as early as 1936 by Tonks \cite{Tonks}. Later in the
1960's, Tonk's model was extended to the more general Lieb-Liniger
model by allowing the strength of particles' contact interaction
($\delta$ function interaction) to be arbitrary. The exact
solution and the thermodynamics of the Lieb-Liniger model have been discussed extensively
\cite{1960-1,1960-2,1960-3,1960-4}. In the last two decades progress in the field of
ultracold atoms has been signifcant
\cite{Coldatoms1,Coldatoms2,Coldatoms3,Coldatoms4} and
possibilities to realize a $\delta$ function
interaction have been proposed
\cite{Delta-1,Delta-2,Delta-3}. Following these theoretical
proposals, the Tonks gas and Lieb-Liniger gas were observed by
several experimental groups \cite{Exp-1,Exp-2,Exp-3}. Due to these
advances, the properties of interacting particles in one dimension
have gained renewed interest \cite{1DAtoms-1,1DAtoms-2}. Beyond
interacting particles in 1D free space, the properties of interacting
particles in a 1D harmonic trap are also of great interest, both in
theory
\cite{1DHamoTrap-T1,1DHamoTrap-T2,1DHamoTrap-T3,1DHamoTrap-T4,1DHamoTrap-T5,1DHamoTrap-T6,1DHamoTrap-T7,1DHamoTrap-T8}
and in experiments \cite{1DHamoTrap-E1,1DAtoms-1,1DAtoms-2}.

Ultracold atoms under periodic  driving are particularly useful
and interesting since periodic driving allows for the generation
of artificial gauge fields
\cite{gauge1,gauge2,gauge3,gaugeM,gauge4,gauge5}. Similar to
Bloch's theorem for spacial periodicity, the Floquet theory
\cite{Floquet1,Floquet2,Floquet3,Floquet4,Floquet5} can be used to
treat time-periodic quantum systems. The Floquet method transforms
a periodic \textit{time-dependent} Schr\"odinger euqation into an
eigenvalue problem of a \textit{time-independent} Floquet
Hamiltonian, which is more accessible to a theoretical treatment.
An intriguing fact of a Floquet Hamiltonian is that it can be used
to simulate time-independent Hamiltonians that are difficult to
access otherwise. Additonally the Floquet Hamiltonian allows for
the study of novel phenomena going beyond equilibrium physics
\cite{FTI,FTI1,FTI2,FTI3,Non_Temp_M,Driven2D,PSL,PSC}. In the
classical limit, the Floquet method corresponds to
$Poincar\acute{e}$ \textit{mapping}
\cite{PoincareMap,PoincareMap1}, which means we observe the
particles every fixed time period and mark them on the phase space
according to their instantaneous positions and momenta. The
particles' trajectories on phase space in the discrete time domain
can be considered as \textit{stroboscopic dynamics}. Most previous
works using Floquet theory are based on the single-particle
picture, i.e., neglecting the interaction between particles.
Recent works \cite{Driven1D1,Driven1D2} realized the importance of
particles' interaction in periodically driven systems and started
to develop scattering theory for Floquet states.

\begin{figure}
\centerline{\includegraphics[scale=0.45]{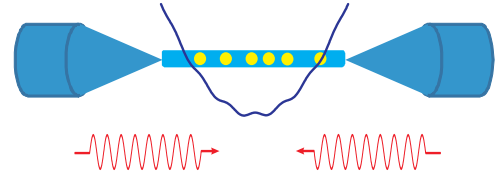}}
\caption{\label{fig Device}{\bf{Sketch of interacting particles in
1D harmonic trap:}} Ultracold atoms (yellow dots) are trapped in
1D harmonic potential. The atoms are further driven by two counter
propagating laser beams (red waves). The intensity of laser beams
is tuned periodically, i.e., $\propto\cos\omega_dt$. The total
potential at a fixed moment (blue rippled curve) is the harmonic
trapping potential plus a cosine function. The interaction
potential between atoms is assumed to be $V(x_i-x_j)$. }
\end{figure}

In this work, we investigate the \textit{classical} dynamics of
interacting identical particles in a 1D harmonic trap in the
presence of various types of interaction potentials, e.g.,
$\delta$ function interaction, Coulomb interaction, hard-core
interaction and Lennard-Jones interaction. The main goal is to
determine how the particles' stroboscopic dynamics is influenced
by their interaction potential $V(x_i-x_j)$. By going to a
rotating frame and using the rotating wave approximation (RWA), we
transform the 1D spatial interaction $V(x_i-x_j)$ to a 2D
interaction in phase space $U(R_{ij})$, which only depends on two
particles's distance in phase space $R_{ij}$. However, there is a
divergence problem in calculating the RWA interaction for, e.g.,
Coulomb interaction. We analyze the origin of divergence and
introduce a renormalization procedure to obtain the correct RWA
interaction. Interestingly, we find a \textit{short-distance}
(e.g., hard-core interaction) real space interaction can generate
a \textit{long-distance} RWA interaction on phase space, which
increases linearly with phase space distance. We justify our
renormalized RWA interaction by simulating the many-body dynamics
numerically and compare the results to our theory.

The article is organized as follows. In section \ref{Model}, we
describe our model of interacting particles in a 1D harmonic trap
under periodic driving. In section \ref{RWA Hamiltonian}, we
introduce the RWA Hamiltonian by transforming to the rotating
frame. We derive the general expression to calculate the RWA
interaction $U(R_{ij})$ for a given real space interaction
$V(x_i-x_j)$. In section \ref{Canonical EOM}, we give the
canonical equations of motion corresponding to both the original
Hamiltonian and the RWA Hamiltonian. To obtain the stroboscopic
dynamics of interacting particles, we also dicuss the
Poincar$\mathrm{\acute{e}}$ mapping method. In section \ref{RWA
Interactions}, we calculate several examples of RWA interactions.
We point out the divergence problem appearing in the case of
Coulomb interaction. Then we introduce the renormalization
procedure to get the correct RWA interaction. We apply our
renormalization method to a more general case of an inverse
power-law interaction potential and the Lennard-Jones interaction
potential. In section \ref{Many-body dynamics}, we investigate the
two-body dynamics and the three-body dynamics with different types
of interactions. We also justify our RWA interaction by showing
the dynamics of eight interacting particles under driving. In
section \ref{Summary and Outlook}, we summarize our work and give
an outlook for future work.



\section{Model}\label{Model}

We consider many identical particles confined in a 1D harmonic
potential and driven by an external driving field  as sketched in
Fig.~\ref{fig Device}. The particles have the same mass $m$ and
harmonic frequency $\omega$. For ultracold atoms, the driving
field can be produced by the interference of two
counter-propagating laser beams \cite{Coldatoms1,Coldatoms2} with
wavelength $a$. The intensity of laser beams is tuned
periodically, i.e., $\propto \cos(\omega_dt)$. We assume the
interaction potential between atoms is $V(x_i-x_j)$. The total
Hamiltonian then is described by
\begin{equation}\label{MainInt.g}
H(t)=\sum_i\Big[\frac{p_i^2}{2}+\frac{x_i^2}{2}+\Lambda\cos(\Omega\
t)\cos(x_i)\Big]+\sum_{i< j}V(x_i-x_j).
\end{equation}
Here, the Hamiltonian has been scaled by the energy unit
$m\omega^2a^2/(4\pi^2)$. The time, coordinate and momentum are
scaled by $\omega^{-1}$, $a/(2\pi)$ and $m\omega a/(2\pi)$
respectively. We also introduced the scaled dimensionless driving
frequency $\Omega\equiv\omega_d/\omega$ and driving strength
$\Lambda\equiv 4\pi^2f/(m\omega^2a^2)$. We define the detuning
parameter via $\delta\equiv1-\Omega/k$, where $k$ is a positive
integer. In this paper, we work in the regime near the resonant
condition, i.e., $|\delta|\ll 1$. Similar Hamiltonians can also be
created using superconducting devices
\cite{superconducting_Devices_H}.

We are interested in the case of weak driving regime $\Lambda \ll
1$, which means we are {\bf not} working on a periodic lattice
model. In our present model, the harmonic trapping potential plays
an important role and the Hamiltonian (\ref{MainInt.g}) does {\bf
not} have spatial periodicity. The basic motion of a particle is
dominated by the global harmonic oscillation with frequency
$\Omega/k$. The driving field and particles' interactions perturb
the phase and amplitude of the global harmonic motion. Thus, the
total motion of a particle can be separated into a fast global
oscillation with frequency $\Omega/k$ and a much slower motion
representing the dynamics of the phase and amplitude of the global
motion
\begin{eqnarray}\label{x_i=P_i}
\left\{
\begin{array}{lll}
x_i(t)&=&P_i(t)\sin\Big(\frac{\Omega}{k}t\Big)+X_i(t)\cos\Big(\frac{\Omega}{k}t\Big)\\
p_i(t)&=&P_i(t)\cos\Big(\frac{\Omega}{k}t\Big)-X_i(t)\sin\Big(\frac{\Omega}{k}t\Big).\\
\end{array}
\right.
\end{eqnarray}
The main task of this paper is to identify the role of interaction
$V(x_i-x_j)$ on the slow dynamics of $X_i(t)$ and $P_i(t)$.

\section{RWA Hamiltonian}\label{RWA Hamiltonian}

We transform the original Hamiltonian (\ref{MainInt.g}) into the
rotating frame with frequency $\Omega/k$ using the generating
function of the second kind
\begin{equation}\label{Int.G2}
G_2(\vec{x},\vec{P},t)=\sum_i\frac{x_iP_i}{\cos(\Omega
t/k)}-\frac{1}{2}x_i^2\tan\Big(\frac{\Omega}{k}t\Big)-\frac{1}{2}P_i^2\tan\Big(\frac{\Omega}{k}t\Big).\nonumber
\end{equation}
Here, $\vec{x}=(x_1,x_2,\cdot\cdot\cdot)$ and
$\vec{P}=(P_1,P_2,\cdot\cdot\cdot)$ represent the assemble
canonical coordinates of all the particles. The corresponding
canonical transformations of coordinates and momenta are given by
$ p_i={\partial G_2(\vec{x},\vec{P},t)}/{\partial x_i},\ \
X_i={\partial G_2(\vec{x},\vec{P},t)}/{\partial P_i}, $ which
result in the transformation (\ref{x_i=P_i}). The canonical
transformation of Hamiltonian itself is given by $H(t)+\partial
G_2/\partial t$. Using the rotating wave approximation (RWA), i.e.,
dropping fast oscillating terms in $H(t)+\partial G_2/\partial
t$, we get the RWA Hamiltonian of all interacting particles in the
rotating frame,
\begin{eqnarray}\label{RWA-g}
g=\sum_i\Big[\frac{1}{2}\delta r_i^2+\Lambda
\cos\Big(\frac{k\pi}{2}\Big)J_k(r_i)\cos(k\theta_i)\Big]+\sum_{i<
j}U(R_{ij}).
\end{eqnarray}
Here, we have defined the vector displacement of the $i$-th
particle in phase space by $r_ie^{i\theta_i}\equiv X_i+iP_i$ and
two particles's \textit{phase space distance} by
$$R_{ij}\equiv|r_ie^{i\theta_i}-r_je^{i\theta_j}|=\sqrt{(X_i-X_j)^2+(P_i-P_j)^2}$$
 The detailed
derivations are given in the Appendix \ref{App. RWA Inter.}. The
RWA approximation is valid for small detuning and weak driving,
i.e., $|\delta|\ll 1$ and $\Lambda \ll 1$.

Normally, the original interaction potential $V(x_i-x_j)$ is only
a function of the two particles' distance $|x_i-x_j|$, which
implies $V(x)=V(-x)$. Given a particular type of interaction
potential $V(x)$, we first define the Fourier transformation of
the potential, i.e., $
V_q=\frac{1}{2\pi}\int_{-\infty}^{+\infty}dx V(x) e^{-iqx}$. Then
the RWA interaction is given by (see more derivation details in
Appendix \ref{App. RWA Inter.})
\begin{eqnarray}\label{RWA-VR}
U(R_{ij})&=&\int_{-\infty}^{+\infty}dq V_q J_0(qR_{ij}),
\end{eqnarray}
where $J_0(\bullet)$ is the Bessel function of zeroth order. Using
the integral representation of the Bessel function, i.e.,
$J_0(x)=\frac{1}{2\pi}\int_{-\pi}^{+\pi}e^{-ix\sin\tau}d\tau$, we
get an alternative form of the RWA interaction as follows
\begin{eqnarray}\label{MainRWA-Ub}
U(R_{ij})=\frac{1}{2\pi}\int_0^{2\pi}V(R_{ij}\sin\tau)d\tau
=\frac{2}{\pi}\int_0^{\frac{\pi}{2}}V(R_{ij}\sin\tau)d\tau.
\end{eqnarray}
We see that $U(R_{ij})$ is in fact the time average of the
interaction energy over the oscillation period under RWA. Since
the RWA interaction $U(R_{ij})$ is defined in the phase space of
the rotating frame and it is only a function of phase space
distance $R_{ij}$, we call it \textit{phase space interaction}.

\section{Canonical Equations of Motion }\label{Canonical EOM}

The time evolution of the original coordinates, $x_i(t)$ and
$p_i(t)$, of a particle are described by the canonical equations
of motion (EOM) according to the original Hamiltonian
(\ref{MainInt.g})
\begin{eqnarray}\label{EOMxp}
\frac{dx_i}{dt}=\frac{\partial H(t)}{\partial p_i},\ \ \ \ \
\frac{dp_i}{dt}=-\frac{\partial H(t)}{\partial x_i}.
\end{eqnarray}
As seen from transformation (\ref{x_i=P_i}), the values of
$X_i(t)$ and $P_i(t)$ can be obtained from the time evolution of
$x_i(t)$ and $p_i(t)$ stroboscopically every time period of
\begin{eqnarray}\label{Dt}
\Delta t=\frac{2k\pi}{\Omega}.
\end{eqnarray}
Here $\Delta t$ is defined as the period of the stroboscopic
dynamics. In this sense, the slow dynamics of $X_i(t)$ and
$P_i(t)$ is defined in the discrete time domain $t=m\Delta t$ with
$m=0,\pm 1,\pm 2,\cdot\cdot\cdot$. This technique is called
$Poincar\acute{e}\ mapping$.

In the rotating frame, the dynamics of $X_i(t)$ and $P_i(t)$ is
described by the canonical EOM according to the RWA Hamiltonian
(\ref{RWA-g}), i.e.,
\begin{eqnarray}\label{MainEOM}
\frac{dX_i}{dt}=\frac{\partial g}{\partial P_i}, \ \ \ \ \ \
\frac{dP_i}{dt}=-\frac{\partial g}{\partial X_i}.
\end{eqnarray}
From the relationship $r_ie^{i\theta_i}= X_i+iP_i$, we have the
explicit form of the canonical EOM (\ref{MainEOM}) as follows (see
the details in Appendix \ref{App. EOM})
\begin{eqnarray}\label{MainEOMRWA}
\left\{
\begin{array}{lll}
\dot{X}_i&=&\delta P_i+\\
&&\Lambda
\cos\Big(\frac{k\pi}{2}\Big)\Big[J'_k(r_i)\cos(k\theta_i)\frac{\partial
r_i}{\partial P_i}-kJ_k(r_i)\sin(k\theta_i)\frac{\partial
\theta_i}{\partial
P_i}\Big]\\
&&+\sum_jU'(R_{ij})\frac{P_i-P_j}{R_{ij}}
\\
\
\\
\dot{P}_i&=&-\  \delta X_i-\\
&&\Lambda
\cos\Big(\frac{k\pi}{2}\Big)\Big[J'_k(r_i)\cos(k\theta_i)\frac{\partial
r_i}{\partial X_i}-kJ_k(r_i)\sin(k\theta_i)\frac{\partial
\theta_i}{\partial
X_i}\Big]\\
&&-\sum_jU'(R_{ij})\frac{X_i-X_j}{R_{ij}} .
\\
\end{array}
\right.
\end{eqnarray}
If we define a complex coordinate via $Z_i\equiv r_ie^{i\theta_i}=
X_i+iP_i$, the EOM (\ref{MainEOMRWA}) can be written in an
alternative form as follows
\begin{eqnarray}\label{EOMRWAZ}
\frac{dZ_i}{dt}&=&-\Big[\Lambda
\cos\Big(\frac{k\pi}{2}\Big)\frac{kJ_k(r_i)}{r^2_i}\sin(k\theta_i)\Big]Z_i\nonumber\\
&&- i\Big[\delta+\Lambda
\cos\Big(\frac{k\pi}{2}\Big)\frac{J'_k(r_i)}{r_i}\cos(k\theta_i)\Big]
Z_i
\nonumber\\
&&-i\sum_jU'(|Z_i-Z_j|)\frac{Z_i-Z_j}{|Z_i-Z_j|} .
\end{eqnarray}
We see that the time evolution of $Z_i(t)$ is determined by three
``forces". The right-hand-side term in the first line of
Eq.(\ref{EOMRWAZ}) is the force parallel to $Z_i$ produced by
driving. The second line of Eq.(\ref{EOMRWAZ}) is the force
perpendicular to $Z_i$ produced by detuning and driving. The third
line of Eq.(\ref{EOMRWAZ}) is the force produced by the
interactions with other particles. The unit vector
$\frac{Z_i-Z_j}{|Z_i-Z_j|}$ represents the direction from $j$-th
particle to $i$-th particle. The interaction strength is
proportional to the derivative of the RWA interaction, i.e.,
$U'(|Z_i-Z_j|)=dU(|Z_i-Z_j|)/d|Z_i-Z_j|$. Different from the
interaction in the laboratory, the imaginary unit $i$ appearing in
this term indicates the direction of this interaction force is
\textit{perpendicular} to the line connecting two particles. Based
on the EOM (\ref{MainEOMRWA}) or (\ref{EOMRWAZ}), we can calculate
all the trajectories of interacting particles on phase space.  The
key of the above EOMs is to determine the explicit form of the RWA
interaction $U(R_{ij})$. Below, we will discuss some examples of
interaction potentials $V(r_i-r_j)$ and calculate their
corresponding RWA interactions $U(R_{ij})$.

\begin{figure}
\centerline{\includegraphics[scale=0.65]{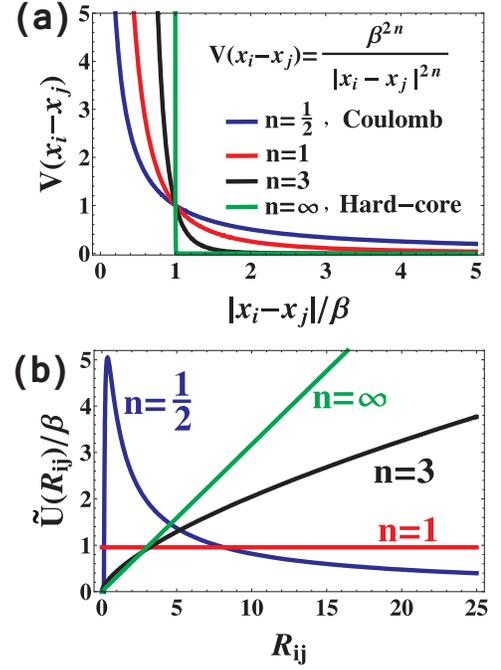}}
\caption{\footnotesize{\textbf{Renormalized RWA Interactions}:
{(a)} Inverse power-law interaction potentials and the example
plots for $n=1/2$ (Coulomb interaction), $n=1$, $n=3$ and
$n=\infty$ (Hard-core interaction).} {(b)} Renormalized RWA
interaction potentials corresponding to the inverse power-law
potentials shown in figure (a). Parameter: $\beta=0.01$.
}\label{fig RWAU}
\end{figure}

\section{RWA Interactions }\label{RWA Interactions}

\subsection{Examples of interaction potentials}

In this subsection, we calculate the RWA interactions for several
specific interaction potentials $V(x_i-x_j)$, i.e., the $\delta$
function interaction potential, the rectangular interaction
potential, the hard-core interaction and the Coulomb interaction
potential. We point out the problem of divergence in the case of
Coulomb interaction, which can be solved by the renormalization
procedure introduced in the next subsection.

\subsubsection{$\delta$ function interaction potential}

The $\delta$ function interaction (contact interaction) is
used to describe the effective interaction between neutral ultra
cold atoms in quasi 1D confinement
\cite{Delta-1,Delta-2,Delta-3,Exp-1,Exp-2,Exp-3}. We describe the
$\delta$ function by a Lorenz-function in the limiting case of vanishing width, i.e.,
 $$
V(x_i-x_j)=\beta\delta(x_i-x_j)=\lim_{\varepsilon\rightarrow
0}\frac{\beta}{\pi}\frac{\varepsilon}{(x_i-x_j)^2+\varepsilon^2}.$$
We introduce the Lorentz-function here, becuase we will later use
it for the numerical simulation. Here, $\beta$ is the strength of
the  $\delta$ function interaction. The Fourier transformation
coefficient of the above Lorentz function is
$V_k={\beta}e^{-|k|\varepsilon}/(2\pi)$. One can obtain the RWA
interaction from Eq. (\ref{RWA-VR}) or  Eq. (\ref{MainRWA-Ub})
\begin{eqnarray}\label{LorentzV}
U(R_{ij})
&=&\frac{\beta}{\pi}\frac{1}{\sqrt{R^2_{ij}+\varepsilon^2}}.
\end{eqnarray}
In the limit of $\varepsilon\rightarrow 0$, we have the RWA
interaction
\begin{eqnarray}\label{deltaV}
U(R_{ij})=\frac{\beta}{\pi R_{ij}}.
\end{eqnarray}
It is interesting to note that the short-distance $\delta$
interaction potential produces an effective long-distance
Coulomb-like interaction as function of phase space distance
$R_{ij}$.

\subsubsection{Rectangular interaction potential }

The $\delta$ function interaction is a point-like interaction with
zero interaction range. Now we allow that the interaction has a
finite range and define a rectangular interaction potential
\[ \eta\  \mathrm{rect}\Big(\frac{x_i-x_j}{2\beta}\Big) = \left\{ \begin{array}{lll}
         0 & \mbox{if $|x_i-x_j| > \beta$} \ ;\\
         \eta/2 & \mbox{if $|x_i-x_j| =\beta$} \ ;\\
         \eta & \mbox{if $|x_i-x_j| <\beta$}\  .
        \end{array} \right. \]
Here, $\beta$ is the interaction range and $\eta$ is the
interaction strength. Applying formula (\ref{MainRWA-Ub}), we have
(see more details in Appendix \ref{App. Examples})
\[U(R_{ij}) = \left\{ \begin{array}{ll}
         \frac{2\eta }{\pi }\arcsin\Big(\frac{\beta}{R_{ij}}\Big) & \mbox{if $R_{ij} \geq \beta$};\\
         \eta & \mbox{if $R_{ij} <\beta$}.
        \end{array} \right. \]
For $R_{ij} \gg \beta$, we have the long-range asymptotic behavior
\begin{eqnarray}\label{MainRectangularU1}
U(R_{ij})\sim \frac{2\eta\  \beta}{\pi }\frac{1}{R_{ij}},\ \ \ \ \
\mathrm{for}\ \ R_{ij} \gg \beta.
\end{eqnarray}
Again, we have an effective long-distance Coulomb-type RWA
interaction from a short-range real space interaction.

\subsubsection{Hard-core interaction potential}

In the  discussion of the rectangular interaction potential, we
have assumed the potential height $\eta$ is finite. This means
that the two particles can overcome the potential barrier and
particles can pass each other if their relative kinetic energy is
large enough. When $\eta$ is larger than the two particles'
relative kinetic energy, the particles can not overcome the
barrier and thus are rebounded back. If the phase space distance
of two particles is $R_{ij}$, their relative kinetic energy is
given by $E_{kin.}(R_{ij})=R^2_{ij}/4$ (see the discussion above
Eq.(\ref{rcCondition})). For the critical condition
$\eta/2=E_{kin.}(R_{ij})$, Eq.(\ref{MainRectangularU1}) becomes
\begin{eqnarray}\label{RectangularU2}
U(R_{ij})\sim \frac{\beta}{\pi} R_{ij}.
\end{eqnarray}
As the potential barrier $\eta$ continues to increase beyond the
critical value $R^2_{ij}/2$, the RWA interaction
(\ref{RectangularU2}) should keep unchanged since the physics in
the rest frame does not change any more. In the limit of
$\eta\rightarrow\infty$, the interaction potential becomes the
hard-core potential, i.e., $V(x_i-x_j)=\infty$ for
$|x_i-x_j|<\beta$ and $V(x_i-x_j)=0$ for $|x_i-x_j|>\beta$.
Therefore, the formula (\ref{RectangularU2}) is the RWA
interaction for hard-core interaction.

\subsubsection{Coulomb interaction potential}

We now consider the case of Coulomb interaction potential, which
is the dominant interaction for trapped ions
\cite{Iontrap,Iontrap1,Iontrap2,Iontrap3}. We approach the Coulomb
potential by using the following function
$$ V(x_i-x_j)=\frac{\beta}{\sqrt{(x_i-x_j)^2+\varepsilon}}\ \ \ \mathrm{with}
\ \ \ \varepsilon>0.$$  Obviously, in the limit of $\varepsilon
\rightarrow 0$, the interaction $V(x_i-x_j)$ becomes the Coulomb
potential. Applying formula (\ref{RWA-VR}) or (\ref{MainRWA-Ub}),
we obtain (more details provided in Appendix \ref{App. Examples})
\begin{eqnarray}\label{MainNearCoulombPotential}
U(R_{ij})=\frac{2\beta^*}{\pi}\frac{1}{\sqrt{R^2_{ij}+\varepsilon}}.
\end{eqnarray}
Here, we introduced the effective coupling $$\beta^*\equiv\beta
K\Big(\sqrt{\frac{R^2_{ij}}{R^2_{ij}+\varepsilon}}\Big),$$ where
$K(x)$ is the complete elliptic integral of the first kind. Using
the asymptotic approximation for the first complete elliptic
integral \cite{EllipticAsymptotic}, i.e., $ K(ix)\approx
\frac{1}{x}\ln(4x)$ for $x\gg1.$,
Eq.(\ref{MainNearCoulombPotential}) has the following
long-distance asymptotic behavior
\begin{eqnarray}\label{MAINNCP}
U(R_{ij})\approx \frac{2\beta}{\pi R_{ij}}\ln
\Big(\frac{4R_{ij}}{\sqrt{\varepsilon}}\Big),\ \ \ \ \mathrm{for}\
\ \  R_{ij}\gg1 .
\end{eqnarray}
We see that the RWA interaction  (\ref{MAINNCP}) diverges in the
limit of $\varepsilon \rightarrow 0$, which means the RWA
interaction given by Eq. (\ref{RWA-VR}) and Eq.(\ref{MainRWA-Ub})
is \textbf{not} valid for the Coulomb interaction potential. We
will analyze the origin of this divergence and introduce the
renormalization procedure to cancel the divergence.

\subsection{Renormalization procedure}

To find the origin of the divergence, we calculate the RWA
interaction for Coulomb interaction potential using
Eq.(\ref{MainRWA-Ub}) by introducing a small cutoff $\tau_c$ as
following
\begin{eqnarray}\label{CoulombUrc1}
U(R_{ij})=\frac{2\beta}{\pi
R_{ij}}\int_{\tau_c\rightarrow0}^{\pi/2}\frac{1}{\sin\tau}d\tau=\frac{2\beta}{\pi
R_{ij}}\ln\Big(\frac{2}{\tau_c}\Big)\Big|_{\tau_c\rightarrow0}.
\end{eqnarray}
We see that $U(R_{ij})$ diverges in the limit of
$\tau_c\rightarrow 0$. Thus, the divergence comes from the
integral contribution inside the small interval $[0,\tau_c]$. For
a given potential $V(r)$ with $r=|x_i-x_j|$, we can estimate the
integral inside the interval $[0,\tau_c]$ by
\begin{eqnarray}\label{shorteffect}
U^{\tau_c}\equiv\frac{2}{\pi}\int_{0}^{\tau_c}V(R_{ij}\sin\tau)d\tau\approx\frac{2}{\pi}\int_{0}^{\tau_c}V(R_{ij}\tau)d\tau.
\end{eqnarray}
If $U^{\tau_c}$ is finite, the potential $V(r)$ is well-behaved
inside the small distance $r<\tau_cR_{ij}$. However, for the
Coulomb potential $V(r)\propto1/r$, $U^{\tau_c}$ is divergent. To
obtain a finite meaningful $U(R_{ij})$, we subtract this
divergence \textit{by hand} and get the following renormalized RWA
interaction
\begin{eqnarray}\label{correctedU}
\tilde{U}(R_{ij})\equiv
U(R_{ij})-U^{\tau_c}=\frac{2}{\pi}\int_{\tau_c}^{\frac{\pi}{2}}V(R_{ij}\sin\tau)d\tau.
\end{eqnarray}
Equivalently, we we introduce the small cutoff $\tau_c$ to remove
the divergence.

The detailed behavior of interaction potential $V(r)$ during the
collision process is crucial. In the real physical process, if
$V(r)\rightarrow\infty$ when $r\rightarrow 0$, the two particles
can never touch each other since the two particles can not
overcome the potential barrier. The smallest distance $r_c$ they
can approach during collision depends on their relative kinetic
energy. If the relative kinetic energy of two particles is much
larger than their interaction energy, the collision distance is
very short, i.e., $r_c\ll 1$. When the two particles are far away,
the interaction energy can be neglect and their kinetic energy in
the center-of-mass frame is calculated
$E_{kin.}(R_{ij})=\frac{1}{2}\Big(\frac{R_{ij}}{2}\Big)^2+\frac{1}{2}\Big(\frac{R_{ij}}{2}\Big)^2=\frac{1}{4}
R_{ij}^2$. Here, the kinetic energy of one particle with respect
to the center of mass is given by
$\frac{1}{2}\Big(\frac{R_{ij}}{2}\Big)^2$. We can calculate $r_c$
according to the energy conservation law in the center-of-mass
frame
\begin{eqnarray}\label{rcCondition}
V(r_c)=E_{kin.}(R_{ij})=\frac{1}{4} R_{ij}^2.
\end{eqnarray}

It is important to note here that $\tilde{U}(R_{ij})$ is in fact
a weighted time average of the interaction in the laboratory
frame as revealed by Eq. (\ref{MainRWA-Ub}). The result is that
two particles which collide during an oscillation period, already
slow down as the they get closer. Therefore, we connect $\tau_c$
and $r_c$ by
\begin{eqnarray}\label{collisionFactor}
 r_c= \gamma R_{ij}\sin \tau_c\approx \gamma R_{ij} \tau_c
\end{eqnarray}
with the important \textit{collision factor} $\gamma \gtrsim 1$,
which is bigger than one, because of the extra interaction energy
the particles accumulate as they move closer to each other and
start to slow down. Therefore we integrate to a radius that can be
smaller than $r_c$. The collision factor $\gamma$ is
phenomenologically introduced here, but we will calculate $\gamma$
for the interaction potentials discussed in this paper. Combining
Eq.s (\ref{correctedU}), (\ref{rcCondition}) and
(\ref{collisionFactor}), we can obtain the explicit form of
renormalized RWA interaction $\tilde{U}(R_{ij})$.


For example, the collision $r_c$ of Coulomb interaction potential
is calculated from Eq. (\ref{rcCondition})
\begin{eqnarray}\label{}
\frac{\beta}{r_c}=\frac{1}{4} R^2_{ij}\ \ \ \Rightarrow \ \ \
r_c=\frac{4\beta}{ R^2_{ij}}.
\end{eqnarray}
Thus, the cutoff $\tau_c$ is given by Eq.(\ref{collisionFactor}),
i.e., $\tau_c\approx\frac{4\beta}{\gamma R^3_{ij}}$. Therefore,
the renormalized RWA interaction for Coulomb interaction is
\begin{eqnarray}\label{CoulombUrc}
\tilde{U}(R_{ij}) =\frac{2\beta}{\pi R_{ij}}\ln(2/\tau_c)
\equiv\frac{2\beta^*}{\pi R_{ij}}.
\end{eqnarray}
Here, $\beta^*\equiv\beta\ \ln(\beta^{-1}\gamma R^3_{ij}/2)$ is
the renormalized coupling strength. The collision factor for
Coulomb potential is $\gamma=e^2$, which is to be calculated in
the next section.

\subsection{Inverse power-law interaction potential }

We now discuss a more general interaction potential form, i.e.,
the inverse power-law interaction potential as plotted in
Fig.~\ref{fig RWAU}(a)
\begin{eqnarray}\label{MainNegative Power law Potential}
V(x_i-x_j)=\frac{\beta^{2n}}{|x_i-x_j|^{2n}}
\end{eqnarray}
We restrict ourselves to integers and half integers $n\geq 1/2$.
If $n=1/2$, the potential $V(x_i-x_j)$ has the form of the Coulomb
potential. If $n\rightarrow \infty$, the potential
$V(x_i-x_j)={\beta^{2n}}/{|x_i-x_j|^{2n}}$ becomes the hard-core
potential with a radius $\beta$. By applying Eq.
(\ref{MainRWA-Ub}), we obtain the RWA interaction
\begin{eqnarray}\label{MainpowerlawV2}
U(R_{ij}) &=&\frac{2\beta}{\pi
R^{2n}_{ij}}\int_{\varepsilon\rightarrow0}^{\pi/2}\frac{1}{\sin^{2n}\tau}d\tau\nonumber\\
&=&\frac{2\beta}{\pi R^{2n}_{ij}}\frac{1}{1-2n}\Big[\
_2F_1(\frac{1}{2},\frac{1}{2}-n;\frac{3}{2}-n;1)\nonumber\\
&&-\varepsilon^{1-2n}\
_2F_1(\frac{1}{2},\frac{1}{2}-n;\frac{3}{2}-n;\varepsilon^2)\Big|_{\varepsilon\rightarrow
0}\Big].
\end{eqnarray}
Due to the term $\varepsilon^{1-2n}$, the above integral diverges
in the limit of $\varepsilon\rightarrow0$ for $n> 1/2$. Below, we
will renormalize $U(R_{ij})$ for integers $n\geq1$, half integers
$n\geq 3/2$ and $n=1/2$ (Coulomb potential) respectively.

\subsubsection{Integers $n\geq1$ }

For integers $n\geq 1$, we use the properties of special functions
$_2F_1(\frac{1}{2},\frac{1}{2}-n;\frac{3}{2}-n;1)=0$ and
$_2F_1(\frac{1}{2},\frac{1}{2}-n;\frac{3}{2}-n;0)=1$. The
renormalized RWA interaction can be obtained from
Eq.(\ref{MainpowerlawV2}) by taking $\varepsilon=\tau_c$
\begin{eqnarray}\label{correctedpowerlawU}
\tilde{U}(R_{ij})=\frac{2\beta^{2n}}{\pi
R^{2n}_{ij}}\frac{1}{2n-1}\tau_c^{1-2n}.
\end{eqnarray}
We use the Eq. (\ref{rcCondition}) to determine the collision
distance $r_c$
\begin{eqnarray}\label{}
\frac{\beta^{2n}}{r^{2n}_c}=\frac{1}{4} R^2_{ij}\ \ \ \Rightarrow
\ \ \ r_c={2^{\frac{1}{n}}}\beta{ R^{-\frac{1}{n}}_{ij}}.
\end{eqnarray}
The truncation $\tau_c$ is given by $\tau_c={2^{\frac{1}{n}}
\beta\gamma^{-1}}{ R^{-1-\frac{1}{n}}_{ij}}$. Plugging $\tau_c$
into Eq.(\ref{correctedpowerlawU}), we get the explicit form of
renormalized RWA interaction
\begin{eqnarray}\label{correctedpowerlawU1}
\tilde{U}(R_{ij})=\frac{2\beta\gamma^{2n-1}4^{\frac{1}{2n}-1}}{\pi(2n-1)}R^{1-\frac{1}{n}}_{ij}.
\end{eqnarray}
The collision factor $\gamma$ will be determined later.

\subsubsection{Half-integers $n\geq 3/2$ }

For half integers $n=k+1/2$ with $k\geq 1$, we have the following
divergence property of Eq. (\ref{MainpowerlawV2})
$$_2F_1(\frac{1}{2},\frac{1}{2}-n;\frac{3}{2}-n;1)=-\sqrt{\pi}\
\frac{\Gamma(k+1/2)}{\Gamma(k)}\frac{\sin(k\pi+\pi/2)}{\cos(k\pi+\pi/2)}\rightarrow\infty.$$
 The half integers can be
approached by taking $n=k+1/2+\epsilon$ with
$\epsilon\rightarrow0$. Thus, we have the asymptotic behavior of
the  divergence above,
\begin{eqnarray}\label{MainAsymptoticA}
_2F_1(\frac{1}{2},\frac{1}{2}-n;\frac{3}{2}-n;1)
\rightarrow\frac{1}{\epsilon}\frac{\Gamma(k+1/2)}{\sqrt{\pi}\
\Gamma(k)}.
\end{eqnarray}
Different from the case of integers $n\geq1$, where we have
neglected the zero function
$_2F_1(\frac{1}{2},\frac{1}{2}-n;\frac{3}{2}-n;1)$ in
Eq.(\ref{MainpowerlawV2}), the divergence (\ref{MainAsymptoticA})
appears in the function
$_2F_1(\frac{1}{2},\frac{1}{2}-n;\frac{3}{2}-n;1)$ for
half-integers $n\geq 3/2$.  It seems that the RWA interaction
(\ref{correctedpowerlawU1}) is not valid for the case of half
integers $n\geq 3/2$. However, we show that this divergence is
artificial and is cancelled by another divergence in the function
of $_2F_1(\frac{1}{2},\frac{1}{2}-n;\frac{3}{2}-n;\varepsilon^2)$.
To reveal this, we write the function in Taylor's series (see
identity (\ref{2F1derivatives}) in Appendix \ref{App. Invers
P.L.})
\begin{eqnarray}\label{}
_2F_1(\frac{1}{2},\frac{1}{2}-n;\frac{3}{2}-n;\varepsilon^2)=\sum_{m=0}^{\infty}\frac{(1/2)_m(-k-\epsilon)_m}{k!(1-k-\epsilon)_m}\varepsilon^{2m}.
\end{eqnarray}
The coefficient for $m=k$ in the limit of $\epsilon\rightarrow 0$
is
\begin{eqnarray}\label{AsymptoticB}
\frac{(1/2)_k(-k-\epsilon)_k}{(1-k-\epsilon)_k}\frac{1}{k!}&=&\frac{1}{\epsilon}\frac{\Gamma(k+1/2)}{\sqrt{\pi}\
\Gamma(k)}.
\end{eqnarray}
This coefficient is divergent as $\epsilon\rightarrow0$ and cancel
the divergence of (\ref{MainAsymptoticA}). Therefore, the RWA
interaction (\ref{correctedpowerlawU1}) is also valid for half
integers $n=k+1/2$ with $k\geq1$.

\begin{figure}
\centerline{\includegraphics[scale=1.0]{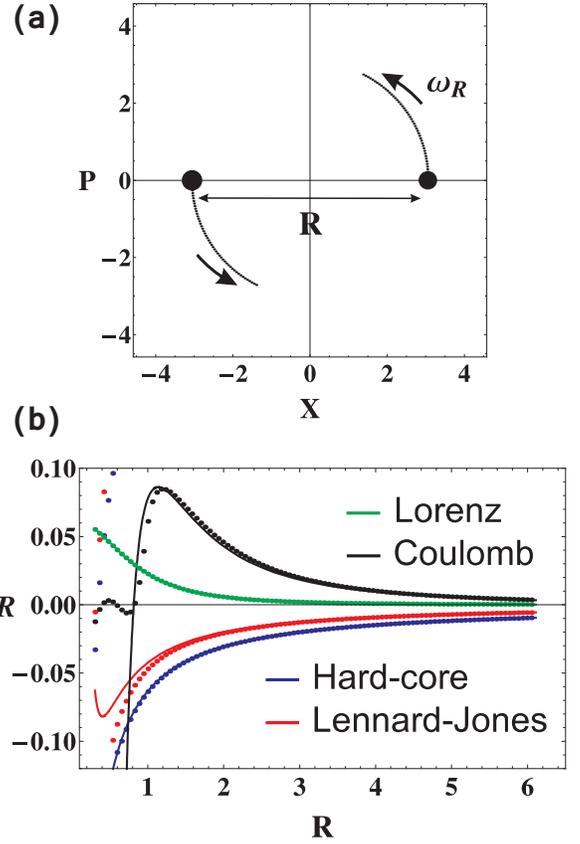}} \caption{
\footnotesize{\textbf{Two-body dynamics for different
interactions:} (a) In the presence of interaction, the two
particles start to rotate at a frequency $\omega_R$, which depends
on their phase space distance $R$. (b) Rotating frequency
$\omega_R$ as function of $R$. The dots represent the data from
Poincar$\acute{e}$ mapping while the solid lines are given by
Eq.(\ref{wP}). Different colors represent different interactions
as indicated on the plot. Interaction potential parameters:
$\beta=0.1, \ \varepsilon=1$ for Lorenz interaction; $\beta=0.1$
for Coulomb interaction; $\epsilon=0.01,\ \sigma=0.1$ for
Lennard-Jones interaction; Hard-core interaction is modelled by
the inverse power-law interaction with $\beta=0.1,\ n=20$. } }
\label{fig 2body}
\end{figure}

\subsubsection{Coulomb potential $n=1/2$ }

Assuming $n=1/2+\epsilon/2$, the power-law interaction potential
(\ref{MainNegative Power law Potential}) becomes
$V(x_i-x_j)=\beta|x_i-x_j|^{-1-\epsilon}$, which goes to the
Coulomb interaction in the limit of $\epsilon\rightarrow0$. Using
the asymptotic property, $
_2F_1(\frac{1}{2},\frac{1}{2}-n;\frac{3}{2}-n;1) \approx
1-\epsilon\ln2$ (see the proof in Appendix \ref{App. Invers
P.L.}), we have from Eq.(\ref{MainpowerlawV2})
\begin{eqnarray}\label{powerlawV3}
U(R_{ij}) =\frac{2\beta}{\pi
R_{ij}}\ln\Big(\frac{2}{\tau_c}\Big)\Big|_{\epsilon=0,\tau_c \ll
1}.
\end{eqnarray}
Compared to formula (\ref{CoulombUrc}), it is just the case of
Coulomb potential (more details are provided in the Appendix
\ref{App. Invers P.L.}).

\subsubsection{Collision factor $\gamma$ }

Now we discuss how to determine the collision factor $\gamma$ for
inverse power-law potential. In principle, the collision factor
$\gamma$ is phenomenologically introduced in
Eq.(\ref{collisionFactor}). Here, we determine it using the
correspondence conditions. From the expression (\ref{MainNegative
Power law Potential}) we see that the inverse power-law potential
approaches the hard-core potential in the limit of
$n\rightarrow\infty$
\[   \frac{\beta^{2n}}{(x_i-x_j)^{2n}} \rightarrow \left\{ \begin{array}{lll}
         \infty & \mbox{if $|x_i-x_j| < \beta$};\\
          0 & \mbox{if $|x_i-x_j| > \beta$}.
        \end{array} \right. \]
Comparing the RWA interaction (\ref{correctedpowerlawU1}) to
Eq.(\ref{RectangularU2}), we get the first correspondence
condition
\begin{eqnarray}
\gamma^{2n-1}/(2n-1)\rightarrow 2\ \ \ \mathrm{for} \ \ \
n\rightarrow\infty.
\end{eqnarray}
The simplest assumption is $\gamma^{2n-1}=4n+c$, where $c$ is a
free parameter to be further determined. Thus the collision factor
takes the form of $\gamma=(4n+c)^{1/(2n-1)}$. This form needs to
be valid for the Coulomb potential, i.e., $n=1/2$. By writing
$n=1/2+\varepsilon/4$ with $\varepsilon\rightarrow 0$, we have
$$\gamma=(2+c+\varepsilon)^{2/\varepsilon}\rightarrow\Big(2+c\Big)^{{2}/{\varepsilon}}e^{2(2+c)}\ \ \
\mathrm{for} \ \ \ \varepsilon\rightarrow0.$$ The parameter $c$
can only take the value of $-1$ to get a meaningful result.
Otherwise, the prefactor $(2+c)^{{2}/{\varepsilon}}\rightarrow
(2+c)^\infty$ takes either zero or infinity, both of which are unphysical.
Finally, we get the expression for the collision factor
$\gamma$ for the inverse power-law interaction potential
\begin{eqnarray}\label{CollisionFactorforPowerPotential}
\gamma=(4n-1)^{\frac{1}{2n-1}}.
\end{eqnarray}
The collision factor becomes $\gamma=e^2$ for Coulomb interaction
($n=1/2$) and $\gamma=1$ for hard-core interaction
($n\rightarrow\infty$).

\begin{figure*}
\centerline{\includegraphics[scale=0.7]{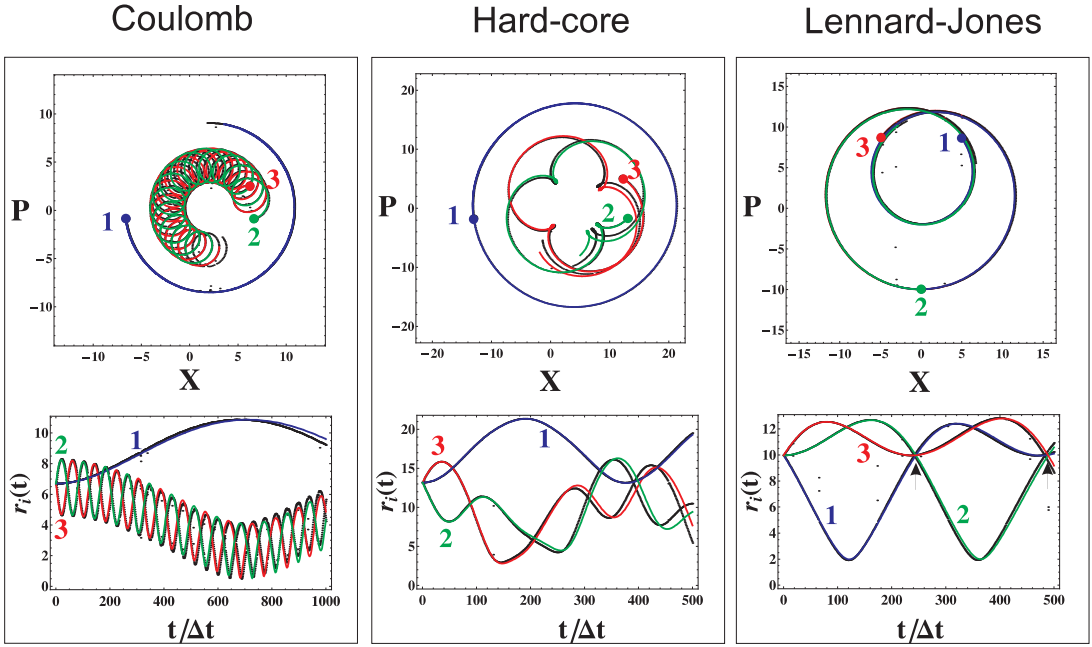}} \caption{
\footnotesize{\textbf{Three-body dynamics for different
interactions:} (Left column) The upper figure shows the
trajectories of three particles in the presence of Coulomb
interaction with $\beta=0.1$. The colored dots represent the
initial conditions and the colored solid lines represent the
dynamics from RWA EOM (\ref{MainEOMRWA}). The black dotted lines
are the data from EOM (\ref{EOMxp}) combined with Poincare
mapping. The lower figure shows the time evolution of
$r_i(t)=\sqrt{X_i^2(t)+P_i^2(t)}, \ i=1,2,3$ obtained from RWA EOM
(\ref{MainEOMRWA}) (solid colored lines) and the
Poincar$\acute{e}$ mapping (dotted black lines). (Middle column)
The middle two figures show the three-body dynamics for hard-core
interaction with $\beta=0.1$}. (Right column) The right two
figures show the three-body dynamics for Lennard-Jones interaction
with parameters $\epsilon=0.01$ and $\sigma=0.1$. } \label{fig
3body}
\end{figure*}

\subsubsection{Summary}
We summarize our results of RWA interaction for the inverse
power-law potential as following
\begin{eqnarray}\label{}
\tilde{ U}(R_{ij}) = \left\{ \begin{array}{lll} \frac{2\beta}{\pi
R_{ij}}\ln(\beta^{-1}\gamma R^3_{ij}/2),\ \  & \mbox{
         for \
         $n=\frac{1}{2}$}\\
         \frac{2\beta\gamma^{2n-1}4^{\frac{1}{2n}-1}}{\pi(2n-1)}R^{1-\frac{1}{n}}_{ij},\ \  & \mbox{ for \  $ n =1, \frac{3}{2},2,\frac{5}{2},\cdot\cdot\cdot$}\\
         \frac{\beta}{\pi}R_{ij},\ \  & \mbox{
         for \
         $n\rightarrow\infty$}.
        \end{array} \right.
\end{eqnarray}
The collision factor is given by $\gamma=(4n-1)^{\frac{1}{2n-1}}$.
We see that the renormalized RWA interaction for the Coulomb
potential ($n=1/2$) still keeps the form of Coulomb's law, up to
logarithmic corrections. We show the behaviors of
$\tilde{U}(R_{ij})$ for several cases in Fig.~\ref{fig RWAU}(b).
For every interaction potential with $n>1$, $\tilde{U}(R_{ij})$
actually grows with $R_{ij}$. It is also interesting to note that
for the case of $n=1$, the corresponding $\tilde{U}(R_{ij})$ is a
constant, which means there is no effective interaction in the
slow dynamics of $X_i(t)$ and $P_i(t)$.

\subsubsection{Lennard-Jones interaction potential}

Another general choice to describe the interaction between two
noble atoms or molecules is the Lennard-Jones interaction
potential
\begin{eqnarray}\label{LJPotential}
V(x_i-x_j)=4\epsilon\Big(
\frac{\sigma^{2m}}{|x_i-x_j|^{2m}}-\frac{\sigma^{m}}{|x_i-x_j|^{m}}
\Big),
\end{eqnarray}
where $\epsilon$ defines the interaction strength, $\sigma$
defines the interaction range and the parameter $m$ is usually
taken $m=6$ in the study. By introducing the cutoff $\tau_c$, we
obtain the RWA interaction from Eq.(\ref{correctedU})
\begin{eqnarray}\label{LJRWA}
\tilde{U}(R_{ij})=\frac{8\epsilon}{\pi}\Big(
\frac{\sigma^{2m}}{R_{ij}^{2m}}\frac{1}{2m-1}\tau_c^{1-2m}-\frac{\sigma^{2m}}{R_{ij}^{m}}\frac{1}{m-1}\tau_c^{1-m}
\Big).
\end{eqnarray}
Then from Eq.(\ref{rcCondition}) and Eq.(\ref{collisionFactor}) we
can calculate the cutoff as following
\begin{eqnarray}\label{LJcutoff}
\tau_c=\frac{\sigma}{\gamma R_{ij}}\Big(\  \frac{1}{2}
+\frac{1}{2}\sqrt{1 + \frac{1}{4\epsilon}R_{ij}^2}  \
\Big)^{-{1}/{m}}.
\end{eqnarray}
As see from Eq. (\ref{LJPotential}), the Lennard-Jones potential
is composed of two inverse power-law potential with exponents $2m$
and $m$ respectively. During the collision in the range of small
distance $|x_i-x_j|<\sigma$, we have
$\frac{\sigma^{2m}}{|x_i-x_j|^{2m}}\gg
\frac{\sigma^{m}}{|x_i-x_j|^{m}}$. Therefore, the term
$\frac{\sigma^{2m}}{|x_i-x_j|^{2m}}$in Eq. (\ref{LJPotential}) is
dominant during the collision. The collision factor of the
Lennard-Jones potential can be calculated from
Eq.(\ref{CollisionFactorforPowerPotential}) by choosing $n=m$,
i.e., $\gamma=(4m-1)^{\frac{1}{2m-1}}$.

\begin{figure*}
\centerline{\includegraphics[scale=1.0]{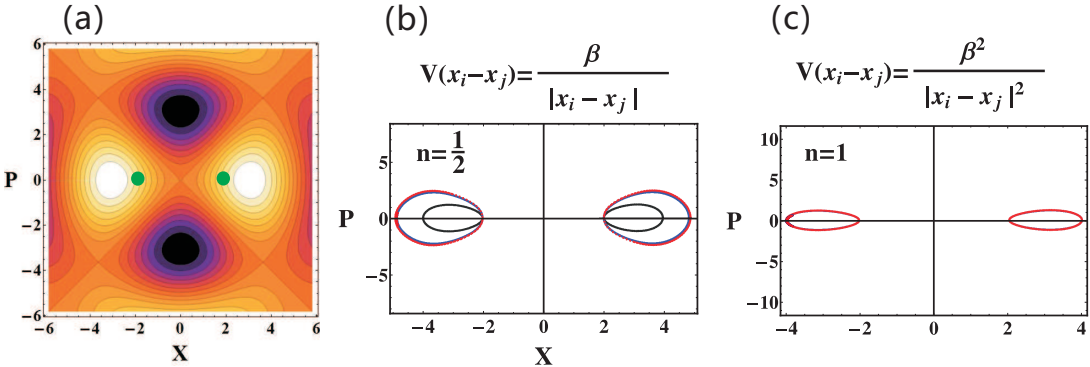}}
\caption{ \footnotesize{\textbf{Two-body dynamics under driving:}
{(a)} Contour plot of single-particle Hamiltonian function (\ref{gi}) in phase space and the
initial conditions of two particles (green dots). {(b)} The
trajectories of two particles in phase space without interaction
(black, $\beta=0$) and with Coulomb interaction (red and blue,
$\beta=0.1$). The black and red trajectories are obtained from the
time evolution based on the original EOM (\ref{EOMxp}) combined
with the technique of $Poincar\acute{e}\ mapping$. The blue
trajectories are obtained from the time evolution based the RWA
EOM (\ref{MainEOMRWA}). Driving parameter: $\Lambda=0.1$. {(c)} The
trajectories of two particles in phase space with interaction
potential $V(x_i-x_j)=\beta^2/(x_i-x_j)^2$. All the trajectories
with interaction (red and blue, $\beta=0.1$ ) and without
interaction (black, $\beta=0$) completely overlap each other.
Driving parameter: $\Lambda=0.1$.}} \label{fig Driving2body}
\end{figure*}

\section{Many-body dynamics}\label{Many-body dynamics}

\subsection{Two-body dynamics}

To justify our RWA interaction, we consider the two-body dynamics
with resonant condition $\delta=0$ and zero driving limit
$\Lambda\rightarrow0$. In Fig.~\ref{fig 2body}(a), we show the
trajectories of two particles in phase space with symmetric
initial conditions, i.e., $(X_1(0)=\frac{1}{2}R,P_1(0)=0)$ and
$(X_2(0)=-\frac{1}{2}R,P_2(0)=\pi)$. Without interaction, both
particles are performing independent harmonic oscillations with
time period $\Delta t=2\pi$. Therefore, the positions of two
particle in  phase space are fixed points as shown by two black
dots in Fig.~\ref{fig 2body}(a). In the presence of interaction,
the two particles start to rotate with a frequency $\omega_R$ as
shown in the same figure. The rotating frequency $\omega_R$
depends on phase space distance $R$ of the two particles. In this
simple case, the motions of the two particles are symmetric. The
dynamics of either particle, e.g., the first particle, is given by
$X_1(t)=\frac{1}{2}R\cos(\omega_R t),\
P_1(t)=\frac{1}{2}R\sin(\omega_R t)$. The rotating frequency
$\omega_R$ can be determined from the EOM (\ref{MainEOMRWA})
\begin{eqnarray}\label{wP}
\omega_R=-\frac{2}{R}\frac{d\tilde{U}(R)}{dR}.
\end{eqnarray}
Here, $\tilde{U}(R)$ is the renormalized RWA interaction.

The two-body rotation frequency $\omega_R$ can also be obtained
from Poincar$\acute{e}$ mapping. We simulate the dynamics of two
particles ($x_i(t),p_i(t)$) based on the EOM (\ref{EOMxp}) and
take their values stroboscopically every time period $\Delta
t=2\pi$. This gives the trajectories of two particles in phase
space as shown by the two dotted lines in Fig.~\ref{fig 2body}(a).
In Fig.~\ref{fig 2body}(b), we compared the numerically extracted
$\omega_R$ to the analytical expression (\ref{wP}) for different
types of interactions, i.e., Lorenz interaction, Coulomb
interaction, hard-core interaction and Lennard-Jones interaction.
We see that the result (\ref{wP}) is very good for large $R$ and
breaks down for small $R$. From Eq.(\ref{collisionFactor}), we see
the relative kinetic energy of two particles is $R^2/4$.
Therefore, we conclude that \textit{the frequency $\omega_R$ as
function of $R$ given by Eq.(\ref{wP}) is valid when the
interaction energy is much smaller than the relative kinetic
energy of two particles, i.e., the cutoff given by
Eq.(\ref{collisionFactor}) satisfies $r_c\ll 1$}.

\begin{figure*}
\centerline{\includegraphics[scale=1.4]{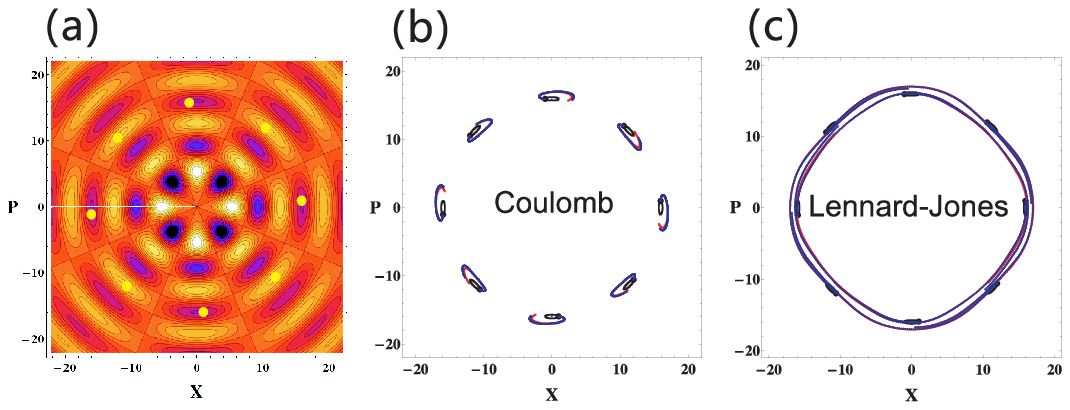}}
\caption{ \footnotesize{\textbf{Eight-body dynamics under
driving:} {(a)} Contour plot of single-particle Hamiltonian
function (\ref{gi}) in phase space with $\delta=0$, $\Lambda=0.1$
and $k=4$. The yellow dots indicate the initial conditions of
eight particles. {(b)} The trajectories of eight particles in
phase space without interaction (black) and with Coulomb
interaction (red and blue, $\beta=0.1$). Driving parameter:
$\Lambda=0.1$. {(c)} The trajectories of eight particles in phase
space with Lennard-Jones interaction (red and blue, $\sigma=0.1$
and $\epsilon=0.01$). Driving parameter: $\Lambda=0.1$. }}
\label{fig Driving8body}
\end{figure*}

\subsection{Three-body dynamics}

As we discussed above, while the two-body dynamics can be solved
analytically, the three-body dynamics can not be obtained
analytically from the RWA EOM (\ref{MainEOMRWA}) in general. Thus,
we solve the three-body problem via numerical simulation. In
Fig.~\ref{fig 3body}, we compare the trajectories based on the
original EOM (\ref{EOMxp}) with the trajectories given by RWA EOM
(\ref{MainEOMRWA}). The results for the Coulomb, hard-core and
Lennard-Jones interactions are given in the left, middle and right
columns of Fig.~\ref{fig 3body} respectively. In the upper figure
of each column, the colored dots (blue, green and red) represent
the three particles' initial positions in phase space. The colored
solid lines are the trajectories obtained from RWA EOM
(\ref{MainEOMRWA}). The black dotted lines are the results from
the original EOM (\ref{EOMxp}) combined with Poincar$\acute{e}$
mapping. In the lower figure of each column, we show the time
evolution of $r_i(t)=\sqrt{X_i^2(t)+P_i^2(t)}, \ i=1,2,3$ obtained
from RWA EOM (\ref{MainEOMRWA}) (solid lines) and the
Poincar$\acute{e}$ mapping (dotted lines).

We see that the agreement is good. The discrepancy in the long
time limit comes from the rotating wave approximation we used in
this paper. Depending on the type of interactions and the initial
conditions, the three interacting particles may have complex
trajectories in phase space. In the figures for Coulomb
interaction, particle 2 and particle 3 can be viewed as a two-body
subsystem while their center of mass forms another larger two-body
system with particle 1. The  RWA interaction corresponding to
Coulomb interaction decays with phase space distance. Thus, as
shown in the lower figure of this column, the rotational frequency
of particle 1 around the center of particle 2 and 3 is much slower
than the rotating frequency of particle 2 and 3 around each other.
For the figures for the hard-core interaction, the the
corresponding RWA interaction increases linearly with the phase
space distance. Therefore the interaction between particle 1 and
particle 2 or between particle 1 and particle 3 is stronger than
interaction between particle 2 and particle 3. As a result, the
orbits of particle 2 and particle 3 are not as regular as in the
case of Coulomb interaction due to the strong disturbance by
particle 1. For the Lennard-Jones interaction, we change the
initial conditions of the three particles. From the lower figure,
we see that the three particles exchange their positions in phase
space with a period of about $240 \Delta t$ as indicated by the
arrows. The parameters for the interactions are given in the
caption of Fig.~\ref{fig 3body}.

\subsection{Dynamics under driving}

In the above discussion, we justify the RWA interaction without
consideration of driving field. Now we add the driving term to the
EOM and justify the RWA interaction. In Fig.~\ref{fig
Driving2body} we show the two-body dynamics under driving. We
choose the driving strength $\Lambda=0.1$ in the RWA Hamiltonian
(\ref{RWA-g}).  We further consider the special resonant condition
that the driving frequency is twice of the harmonic frequency in
the RWA Hamiltonian (\ref{RWA-g}), i.e., $g=g_1+g_2+U(R_{12})$.
Here, $g_{i}$ with $i=1,2$ representing the index of two particles
is defined as the single-particle Hamiltonian function
\begin{eqnarray}\label{gi}
g_i=\frac{1}{2}\delta r_i^2+\Lambda \cos\Big(\frac{k}{2}\pi\Big)
J_k(r_i)\cos(k\theta_i).
\end{eqnarray}
In Fig.~\ref{fig Driving2body}(a) we plot the single-particle
Hamiltonian function in phase space with chosen parameters
$\delta=0$ and $k=2$, i.e., $g_i=-\Lambda
J_2(r_i)\cos(2\theta_i)$. Due to the driving field, the system
hosts multiple stable vibrational states with different amplitudes
and phases. We only draw the four lowest vibrational stable states
with different phases (centers of the bright and dark regions).
Now we put two particles (two green dots) near two stable states
with phase $0$ and $\pi$ respectively, and study their motions in
the presence of interaction. In Fig.~\ref{fig Driving2body}(b) and
Fig.~\ref{fig Driving2body}(c), we compare the full numerical
simulations to the RWA dynamcis. The trajectories with black and
red colors are the results from the time evolution based on the
original EOM (\ref{EOMxp}) with consideration of the two
particles' interaction $V(x_i-x_j)$ combined with the technique of
Poincar$\acute{e}$ mapping. The trajectories with blue color are
the results from the time evolution based on the RWA EOM
(\ref{MainEOMRWA}) with the renormalized $\tilde{U}(R_{ij})$. We
see that the results agree with each other. Especially we find,
for the case of $n=1$, the trajectories for $\beta=0$ and
$\beta\neq 0$ overlap completely as shown in Fig.~\ref{fig
Driving2body}(c). This means the inverse power-law potential for
$n=1$ generates no interaction under RWA.

Fig.~\ref{fig Driving8body} shows the dynamics of eight
interacting particles. In Fig.~\ref{fig Driving8body}(a) we plot
the single-particle Hamiltonian function (\ref{gi}) with
parameters $\delta=0$, $\Lambda=0.1$ and $k=4$. The initial
conditions are represented by the eight yellow dots. In
Fig.~\ref{fig Driving8body}(b) and Fig.~\ref{fig Driving8body}(c)
we show the trajectories of eight particles under driving. Without
interaction the eight particles exhibit localized motions as
illustrated by the black curves. In the case of Coulomb
interaction, the trajectories of eight particles are enlarged a
bit as shown in Fig.~\ref{fig Driving8body}(b) by the red orbits
(from Poincar$\acute{e}$ mapping) and blue orbits (from RWA EOM
(\ref{MainEOMRWA})). The RWA interaction corresponding to Coulomb
interactions decays with phase space distance, which can be viewed
as small perturbation when the particle are far away in phase
space. In the case of the Lennard-Jones interaction, the
trajectories of eight particles become global motions as shown in
Fig.~\ref{fig Driving8body}(c). This  reflects the fact that the
RWA interaction corresponding to the Lennard-Jones interaction
increases as the phase space distance increases, which is similar
to the hard-core interaction.

\section{Summary and Outlook}\label{Summary and Outlook}

We investigated the classical dynamics of periodically driven
interacting particles in a 1D harmonic trap. Under RWA, we
transform the real space interaction $V(x_i-x_j)$ into a RWA
interaction in phase space. Particularly, we find an effective
long-distance RWA interaction can be produced by short-distance
real space interactions, e.g., point-like $\delta$ function
interaction and hard-core interaction. The RWA interaction
describes the effect of the interaction on the slowly changing
amplitude and phase of the globally oscillating particles, while
the fast oscillations take on the role of a force carrier, which
allows for interaction over much larger effective distances.

We solved the divergence problem by introducing the
renormalization procedure. For the Coulomb interaction, our
renormalization procedure just eliminates the high energy
collision process, which gives rise to a renormalized strength of
Coulomb interaction. For the hard-core interaction, our
renormalization procedure gives rise to a completely different
long-distance interaction, which increases linearly with phase
space distance. We justified our theory by simulating the
many-body dynamics numerically in the presence of various
interaction potentials like point-like $\delta$ interaction,
Coulomb interaction, hard-core interaction and Lennard-Jones
interaction.

The work in this paper only focuses on the classical dynamics of
interacting 1D particles. The next step is to extend our study to
the quantum regime. An interesting direction is combing the
lattice structure created by the driving \cite{1DHamoTrap-T6}, as
shown in Fig.~\ref{fig Driving8body}(a), together with the
effective RWA interaction. In this way, a Hubbard model can be
simulated in phase space. This will provide another way to study
strongly correlated systems by periodically driving 1D system.

\bigskip

\textbf{Acknowledgements}

L. Guo and M. Liu acknowledge support from the KHYS-Gaststipendium
(Visiting Researcher Scholarship from Karlsruhe House of Young
Scientist, KIT). L. Guo also acknowledges financial support from
Carl-Zeiss Stiftung.

\appendix

\section{Derivation of RWA Interaction}\label{App. RWA Inter.}

The Hamiltonian of many trapped interacting particles under
periodic driving can be written as
\begin{equation}\label{Int.g}
H(t)=\sum_i\Big[\frac{p^2_i}{2}+\frac{x^2_i}{2}+\Lambda\cos(\Omega\
t)\cos(x_i)\Big]+\sum_{i< j}V(x_i-x_j).
\end{equation}
Here, all the quantities are scaled to be dimensionless. We are
working in the regime near the resonant condition, i.e.,
$\Omega\approx k$ with a positive integer $k$. Using the
generating function of the second kind
\begin{equation}\label{Int.G2}
G_2(\vec{x},\vec{P},t)=\sum_i\frac{x_iP_i}{\cos(\Omega
t/k)}-\frac{1}{2}x_i^2\tan\Big(\frac{\Omega}{k}t\Big)-\frac{1}{2}P_i^2\tan\Big(\frac{\Omega}{k}t\Big),
\end{equation}
we transform to the rotating frame with frequency $\Omega/k$.
Here, $\vec{x}=(x_1,x_2,\cdot\cdot\cdot)$ and
$\vec{P}=(P_1,P_2,\cdot\cdot\cdot)$ represent the assemble
canonical coordinates of all the particles. The corresponding
canonical transformations of coordinates and momenta is
\begin{equation}
p_i=\frac{\partial G_2(\vec{x},\vec{P},t)}{\partial x_i},\ \ \
X_i=\frac{\partial G_2(\vec{x},\vec{P},t)}{\partial P_i},
\end{equation}
which results in
\begin{equation}
x_i=P_i\sin\Big(\frac{\Omega}{k}t\Big)+X_i\cos\Big(\frac{\Omega}{k}t\Big),\
\ \
p_i=P_i\cos\Big(\frac{\Omega}{k}t\Big)-X_i\sin\Big(\frac{\Omega}{k}t\Big).
\end{equation}
The canonical transformation of $H(t)$ itself is
\begin{eqnarray}\label{cano.H}
K(t)&\equiv&H(t)+\partial G_2/\partial t\nonumber\\
&=&\sum_i\frac{1}{2}\delta(X_i^2+P_i^2)\nonumber\\
&&+\Lambda\cos(\Omega t)\cos\Big(P_i\sin(\Omega
t/k)+X_i\cos(\Omega t/k)\Big)\nonumber\\
&&+\sum_{i<
j}V\Big(\Delta P_{i,j}\sin(\Omega t/k)+\Delta X_{i,j}\cos(\Omega t/k)\Big)\nonumber\\
&=&\sum_i\frac{1}{2}\delta(X_i^2+P_i^2)\nonumber\\
&&+\frac{\Lambda}{2}\Big(\cos(\Omega t)e^{i[P_i\sin(\Omega
t/k)+X_i\cos(\Omega t/k)]}+h.c.\Big)\nonumber\\
&& +\sum_{i< j}V\Big(\Delta P_{i,j}\sin(\Omega t/k)+\Delta
X_{i,j}\cos(\Omega t/k)\Big).
\end{eqnarray}
Here, we have defined the detuning $\delta\equiv1-\Omega/k$, the
displacement of one particle in phase space
$r_ie^{i\theta_i}\equiv X_i+iP_i$, and the relative displacement
of two particles in phase space $(\Delta X_{ij},\Delta
P_{ij})\equiv(X_i-X_j,P_i-P_j).$ Using the Jacobi-Anger expansion,
$$e^{iz\cos\varphi}=\sum_{-\infty}^{+\infty}i^mJ_m(z)e^{im\varphi},$$
we have
\begin{eqnarray}\label{non-RWAs}
&&\cos(\Omega t)e^{i[P_i\sin(\Omega t/k)+X_i\cos(\Omega t/k)]}\nonumber\\
&=&\cos(\Omega t)e^{ir_i\cos(\Omega t/k-\theta_i)}\nonumber\\
&=&\cos(\Omega t)\sum_{-\infty}^{+\infty}i^mJ_m(r_i)e^{-im\theta_i}e^{im\Omega t/k}\nonumber\\
&=&\frac{1}{2}\Big(e^{i\Omega t}+e^{-i\Omega
t}\Big)\sum_{-\infty}^{+\infty}i^mJ_m(r_i)e^{-im\theta_i}e^{im\Omega
t/k}.
\end{eqnarray}
Under rotating wave approximation(RWA), we drop fast oscillating
terms in Eq.(\ref{non-RWAs}) and obtain
\begin{eqnarray}\label{RWAs1}
&&\cos(\Omega t)e^{i[P_i\sin(\Omega t/k)+X_i\cos(\Omega t/k)]}
\longrightarrow\nonumber\\
&&\frac{1}{2}\Big(i^kJ_k(r_i)e^{-ik\theta_i}+i^{-k}J_{-k}(r_i)e^{ik\theta_i}\Big)=e^{ik\frac{\pi}{2}}J_k(r_i)\cos(k\theta_i).\nonumber\\
\end{eqnarray}
Now we focus on the interacting term in Eq.(\ref{cano.H})
\begin{eqnarray}\label{nonRWA-V}
&&V\Big(\Delta P_{i,j}\sin(\Omega t/k)+\Delta X_{i,j}\cos(\Omega
t/k)\Big)\nonumber\\
&=&V\Big( R_{i,j}\cos(\Omega t/k-\theta_{i,j})\Big)\nonumber\\
&=&\int_{-\infty}^{+\infty}dq V_q
e^{iqR_{i,j}\cos(\Omega t/k-\theta_{i,j})}\nonumber\\
&=&\sum_{m=-\infty}^{+\infty}i^m\int_{-\infty}^{+\infty}dq V_q
J_m(qR_{ij}) e^{im(\Omega t/k-\theta_{i,j})}.
\end{eqnarray}
Here, $R_{ij}e^{i\theta_{ij}}\equiv (\Delta X_{ij},\Delta
P_{ij})$, $V_q$ is the coefficient of Fourier transformation of
potential, i.e., $$V_q=\frac{1}{2\pi}\int_{-\infty}^{+\infty}dx
V(x) e^{-iqx}.$$ In the RWA, we drop all the terms with $m\neq 0$
in Eq.(\ref{nonRWA-V}) and obtain the RWA interaction potential in
phase space
\begin{eqnarray}\label{RWA-V}
U(R_{ij})=\int_{-\infty}^{+\infty}dq V_q J_0(qR_{ij}).
\end{eqnarray}
We then apply the integral representation of Bessel function,
i.e.,
$$J_0(x)=\frac{1}{2\pi}\int_{-\pi}^{+\pi}e^{-ix\sin\tau}d\tau,$$ and
get an alternative form of Eq. (\ref{RWA-V})
\begin{eqnarray}\label{RWA-U}
U(R_{ij})&=&\int_{-\infty}^{+\infty}dq V_q J_0(qR_{ij})\nonumber\\
&=&\int_{-\infty}^{+\infty}dq
V_q\frac{1}{2\pi}\int_{-\pi}^{+\pi}e^{-iqR_{ij}\sin\tau}d\tau\nonumber\\
&=&\frac{1}{2\pi}\int_{-\pi}^{+\pi}V(R_{ij}\sin\tau)d\tau.
\end{eqnarray}
Normally, the potential is only a function of the distance
$|x_i-x_j|$ which implies $V(x)=V(-x)$. Thus we have
\begin{eqnarray}\label{RWA-Ub}
U(R_{ij})
&=&\frac{1}{2\pi}\int_{-\pi}^{+\pi}V(R_{ij}\sin\tau)d\tau
\nonumber\\
&=&\frac{2}{\pi}\int_0^{\frac{\pi}{2}}V(R_{ij}\sin\tau)d\tau.
\end{eqnarray}
Finally, we get the RWA Hamiltonian of all interacting particles
in phase space
\begin{eqnarray}\label{Int.RWAg}
g=\sum_i\Big[\frac{1}{2}\delta r_i^2+\Lambda
\cos\Big(\frac{k\pi}{2}\Big)J_k(r_i)\cos(k\theta_i)\Big]+\sum_{i<
j}U(R_{ij}),\nonumber\\
\end{eqnarray}
where
$R_{ij}=|r_ie^{i\theta_i}-r_je^{i\theta_j}|=\sqrt{(X_i-X_j)^2+(P_i-P_j)^2}$
is the phase space distance.

\section{Canonical equations of motion in the rotating
frame}\label{App. EOM}

The canonical equations of motion given by RWA Hamiltonian
(\ref{Int.RWAg}) in the rotating frame are
\begin{eqnarray}\label{EOM}
\frac{dX_i}{dt}=\frac{\partial g}{\partial P_i}, \ \ \ \ \ \
\frac{dP_i}{dt}=-\frac{\partial g}{\partial X_i}.
\end{eqnarray}
From the relationship
\begin{eqnarray}\label{eom1}
X_i=r_i\cos\theta_i,\ \  P_i=r_i\sin\theta_i \ \ \ \mathrm{and}\ \
\ r_i=\sqrt{X_i^2+P_i^2},
\end{eqnarray}
we have
\begin{eqnarray}\label{eom2}
\frac{\partial \theta_i}{\partial X_i}=\frac{-P_i}{X_i^2+P_i^2}, \
&&  \ \frac{\partial \theta_i}{\partial
P_i}=\frac{X_i}{X_i^2+P_i^2} \nonumber\\ \frac{\partial
r_i}{\partial X_i}=\frac{X_i}{\sqrt{X_i^2+P_i^2}}, \ &&
\frac{\partial r_i}{\partial P_i}=\frac{P_i}{\sqrt{X_i^2+P_i^2}}.
\end{eqnarray}
As a result, the explicit form of canonical equations of motion
are
\begin{eqnarray}\label{EOMRWA}
\frac{dX_i}{dt}&=& \delta P_i\nonumber\\
&&+\Lambda
\cos\Big(\frac{k\pi}{2}\Big)\Big[J'_k(r_i)\cos(n\theta_i)\frac{\partial
r_i}{\partial P_i}-kJ_k(r_i)\sin(k\theta_i)\frac{\partial
\theta_i}{\partial
P_i}\Big]\nonumber\\
&&+\sum_jU'(R_{ij})\frac{P_i-P_j}{R_{ij}},\nonumber\\
\frac{dP_i}{dt}&=&- \delta X_i\nonumber\\
&&-\Lambda
\cos\Big(\frac{k\pi}{2}\Big)\Big[J'_k(r_i)\cos(k\theta_i)\frac{\partial
r_i}{\partial X_i}-kJ_k(r_i)\sin(k\theta_i)\frac{\partial
\theta_i}{\partial
X_i}\Big]\nonumber\\
&&-\sum_jU'(R_{ij})\frac{X_i-X_j}{R_{ij}} .
\end{eqnarray}
Based on EOM (\ref{EOMRWA}), we can calculate the trajectories of
interacting particles in phase space.

\section{Examples of interaction potentials}\label{App. Examples}

\subsection{Rectangular potential }

We define the following short-range interaction potential by a
rectangular function
\[ \eta\  \mathrm{rect}\Big(\frac{x_i-x_j}{2\beta}\Big) = \left\{ \begin{array}{lll}
         0 & \mbox{if $|x_i-x_j| > \beta$};\\
         \eta/2 & \mbox{if $|x_i-x_j| =\beta$};\\
         \eta & \mbox{if $|x_i-x_j| <\beta$}.
        \end{array} \right. \]
Applying formula (\ref{RWA-Ub}), we have
\begin{eqnarray}\label{RectangularU}
U(R_{ij}) &=&\frac{2\eta}{\pi}\int_{0}^{\pi/2}\
\mathrm{rect}\Big(\frac{R_{ij}\sin\tau}{2\beta}\Big)d\tau\nonumber\\
&=&\frac{4\beta\eta}{\pi R_{ij}}\int_{0}^{\frac{R_{ij}}{2\beta}}\
\frac{\mathrm{rect}(x)}{\sqrt{1-(\frac{2\beta}{R_{ij}}x)^2}}dx,
\end{eqnarray}
where $x=\frac{R_{ij}\sin\tau}{2\beta}$. Using the definition of
$\mathrm{rect}(x)$, we have
\[U(R_{ij}) = \left\{ \begin{array}{ll}
         \frac{2\eta }{\pi }\arcsin\Big(\frac{\beta}{R_{ij}}\Big) & \mbox{if $R_{ij} \geq \beta$};\\
         \eta & \mbox{if $R_{ij} <\beta$}.
        \end{array} \right. \]
For $R_{ij} \gg \beta$, we have the long-range asymptotic behavior
\begin{eqnarray}\label{RectangularU1}
U(R_{ij})\sim \frac{2\eta\  \beta}{\pi }\frac{1}{R_{ij}},\ \ \ \ \
\mathrm{for}\ \ R_{ij} \gg \beta.
\end{eqnarray}

\subsection{Coulomb potential}

We approach the Coulomb potential by the following type of
interaction potential
$$ V(x_i-x_j)=\frac{\beta}{\sqrt{(x_i-x_j)^2+\varepsilon}}\ \ \ \mathrm{with}
\ \ \ \varepsilon>0.$$ The interaction $V(x_i-x_j)$ goes to
Coulomb potential in the limit of $\varepsilon \rightarrow 0$.
Applying formula (\ref{RWA-U}), we obtain
\begin{eqnarray}\label{NearCoulombPotential}
U(R_{ij})&=&\frac{\beta}{2\pi}\int_{-\pi}^{+\pi}\frac{1}{\sqrt{R^2_{ij}\sin^2\tau+\varepsilon}}d\tau\nonumber\\
&=&\frac{2\beta}{\pi \sqrt{\varepsilon}}\int_{0}^{\pi/2}\frac{1}{\sqrt{\varepsilon^{-1} R^2_{ij}\sin^2\tau+1}}d\tau\nonumber\\
&=&\frac{2\beta}{\pi \sqrt{\varepsilon}}
K\Big(i\frac{R_{ij}}{\sqrt{\varepsilon}}\Big)\nonumber\\
&\equiv&\frac{2\beta^*}{\pi}\frac{1}{\sqrt{R^2_{ij}+\varepsilon}}.
\end{eqnarray}
Here, we introduced the effective coupling $\beta^*\equiv\beta
K\Big(\sqrt{\frac{R^2_{ij}}{R^2_{ij}+\varepsilon}}\Big)$, where
$K(x)$ is the complete elliptic integral of the first kind with
the property of
$$K(ix)=\frac{1}{\sqrt{1+x^2}}K\Big(\sqrt{\frac{x^2}{x^2+1}}\Big)$$
for a real $x$. We use the asymptotic approximations for the first
complete elliptic integral \cite{EllipticAsymptotic}, i.e., $
K(ix)\approx \frac{1}{x}\ln(4x)$ for $x\gg1.$ Then we have the
long-range asymptotic behavior of potential
(\ref{NearCoulombPotential})
\begin{eqnarray}\label{NCP}
U(R_{ij})\approx \frac{2\beta}{\pi R_{ij}}\ln
\Big(\frac{4R_{ij}}{\sqrt{\varepsilon}}\Big),\ \ \ \ \mathrm{for}\
\ \  R_{ij}\gg1 .
\end{eqnarray}
We see that in the limit of $\varepsilon \rightarrow 0$, where the
real space interaction $V(x_i-x_j)$ goes to Coulomb potential, the
phase space interaction potential (\ref{NCP}) diverges. We will
analyze the physical origin of this divergence and introduce the
renormalization procedure to cancel it in the next section

\section{Inverse power-law potential }\label{App. Invers P.L.}

We assume an interaction potential in the form of inverse
power-law, i.e.,
\begin{eqnarray}\label{Negative Power law Potential}
V(x_i-x_j)=\frac{\beta^{2n}}{|x_i-x_j|^{2n}}
\end{eqnarray}
with integers and half integers $n\geq1/2$. We apply the formula
(\ref{RWA-Ub}) and obtain
\begin{eqnarray}\label{powerlawV2}
U(R_{ij}) &=&\frac{2\beta}{\pi
R^{2n}_{ij}}\int_{0}^{\pi/2}\frac{1}{\sin^{2n}\tau}d\tau\nonumber\\
&=&\frac{2\beta}{\pi
R^{2n}_{ij}}\int_{0}^{1}\frac{1}{t^{2n}\sqrt{1-t^2}}dt\nonumber\\
&=&\frac{2\beta}{\pi R^{2n}_{ij}}\frac{t^{1-2n}}{1-2n}\
_2F_1(\frac{1}{2},\frac{1}{2}-n;\frac{3}{2}-n;t^2)\Big|_{\varepsilon\rightarrow
0}^1\nonumber\\
&=&\frac{2\beta}{\pi R^{2n}_{ij}}\frac{1}{1-2n}\Big[\
_2F_1(\frac{1}{2},\frac{1}{2}-n;\frac{3}{2}-n;1)\nonumber\\
&&-\varepsilon^{1-2n}\
_2F_1(\frac{1}{2},\frac{1}{2}-n;\frac{3}{2}-n;\varepsilon^2)\Big|_{\varepsilon\rightarrow
0}\Big].
\end{eqnarray}
Due to the term $\varepsilon^{1-2n}$, the above integral diverges
in the limit of $\varepsilon\rightarrow0$ for $n> 1/2$. We
renormalize $U(R_{ij})$ for integers $n\geq1$, half integers
$n\geq 3/2$ and $n=1/2$ (Coulomb potential) respectively.

\subsection{Identities of hypergeometric function}

To continue, we should introduce some identities of hypergeometric
function $_2F_1(a,b;c;z)$ and Gamma function $\Gamma(z)$.

\ \\
\textbf{Identity I:} Gauss's theorem\cite{HypergeometricFunctions}
of hypergeometric function $_2F_1(a,b;c;1)$
\begin{eqnarray}\label{Gauss's theorem}
_2F_1(a,b;c;1)=\frac{\Gamma(c)\Gamma(c-a-b)}{\Gamma(c-a)\Gamma(c-b)},\
\ \ for \ \ \ \mathrm{Re}[c]>\mathrm{Re}[a+b].\nonumber\\
\end{eqnarray}
\textbf{Identity II:} derivatives of hypergeometric function
$_2F_1(a,b;c;z)$
\begin{eqnarray}\label{2F1derivatives}
&&\frac{d^m}{dz^m}\ _2F_1(a,b;c;z)=\frac{(a)_m(b)_m}{(c)_m}\
_2F_1(a+m,b+m;c+m;z),\nonumber\\
&& \ \ \mathrm{where} \ \ \
(q)_m\equiv\frac{\Gamma(q+m)}{\Gamma(q)}.
\end{eqnarray}
\textbf{Identity III:} Euler's reflection formula
\cite{ReflectionRelation} of Gamma function
\begin{eqnarray}\label{Euler's reflection formula}
\Gamma(1-z)\Gamma(z)=\frac{\pi}{\sin(\pi z)}.
\end{eqnarray}
\textbf{Identity IV:} Euler's duplication formula
\cite{ReflectionRelation} of Gamma function
\begin{eqnarray}\label{Euler's duplication formula}
\Gamma(z)\Gamma(z+\frac{1}{2})=2^{1-2z}\sqrt{\pi}\Gamma(2z).
\end{eqnarray}
\textbf{Identity V:} Taylor's expansion of Gamma function
\begin{eqnarray}\label{Laurent expansion}
\ln\Gamma(1+z)=-\gamma
z+\sum_{k=1}^{\infty}\frac{\zeta(k)}{k!}(-z)^k,\ \ \ for \ \ \
|z|<1,
\end{eqnarray}
where $\gamma\approx0.5772156649$ is the Euler-Mascheroni constant
and $\zeta(k)$ is the Riemann zeta function at $k$. Combined with
Eq.(\ref{Euler's reflection formula}), we have the following
approximative identity
\begin{eqnarray}\label{Laurent expansion1}
\Gamma(z)\approx\frac{1}{z}\ e^{-\gamma z},\ \ \ for \ \ \
|z|\ll1,
\end{eqnarray}

\subsection{Half-integers $n\geq 3/2$ }

For half integers $n=k+1/2$ with $k\geq 1$, we have the following
divergence in formula (\ref{powerlawV2})
$$_2F_1(\frac{1}{2},\frac{1}{2}-n;\frac{3}{2}-n;1)=-\sqrt{\pi}\
\frac{\Gamma(k+1/2)}{\Gamma(k)}\frac{\sin(k\pi+\pi/2)}{\cos(k\pi+\pi/2)}\rightarrow\infty.$$
It seems the phase space potential (\ref{correctedpowerlawU1}) is
not valid for the case of half integers $n\geq 3/2$. However, we
show this divergence is artificial and is cancelled by another
divergence in formula (\ref{powerlawV2}). The half integers can be
approached by taking $n=k+1/2+\epsilon$ with
$\epsilon\rightarrow0$. Then we have
\begin{eqnarray}\label{AsymptoticA}
_2F_1(\frac{1}{2},\frac{1}{2}-n;\frac{3}{2}-n;1)&=&\sqrt{\pi}\
\frac{\Gamma(k+1/2+\epsilon)}{\Gamma(k+\epsilon)}\frac{\cos(k\pi+\pi\epsilon)}{\sin(k\pi+\pi\epsilon)}\nonumber\\
&\rightarrow&\frac{1}{\epsilon}\frac{\Gamma(k+1/2)}{\sqrt{\pi}\
\Gamma(k)}.
\end{eqnarray}
In fact, there is also a divergent term in the function of
$_2F_1(\frac{1}{2},\frac{1}{2}-n;\frac{3}{2}-n;\varepsilon^2)$. To
reveal it, we write the function
$_2F_1(\frac{1}{2},\frac{1}{2}-n;\frac{3}{2}-n;\varepsilon^2)$ in
Taylor's series using identity (\ref{2F1derivatives})
\begin{eqnarray}\label{}
&&_2F_1(\frac{1}{2},\frac{1}{2}-n;\frac{3}{2}-n;\varepsilon^2)\nonumber\\
&=&\sum_{m=0}^{\infty}\varepsilon^{2m}\frac{(1/2)_m(-k-\epsilon)_m}{k!(1-k-\epsilon)_m}\
_2F_1(\frac{1}{2},\frac{1}{2}-n;\frac{3}{2}-n;0)\nonumber\\
&=&\sum_{m=0}^{\infty}\frac{(1/2)_m(-k-\epsilon)_m}{k!(1-k-\epsilon)_m}\varepsilon^{2m}.
\end{eqnarray}
We calculate the coefficient for $m=k$ in the limit of
$\epsilon\rightarrow 0$
\begin{eqnarray}\label{AsymptoticB}
&&\frac{(1/2)_k(-k-\epsilon)_k}{(1-k-\epsilon)_k}\frac{1}{k!}\nonumber\\
&=&\frac{\Gamma(k+1/2)}{\Gamma(1/2)}\frac{\Gamma(-\epsilon)}{\Gamma(-k-\epsilon)}
\frac{\Gamma(1-k-\epsilon)}{\Gamma(1-\epsilon)}\frac{1}{\Gamma(k+1)}\nonumber\\
&=&\frac{\Gamma(k+1/2)}{\Gamma(1/2)}\frac{\Gamma(-\epsilon)}{(-\epsilon)\Gamma(-\epsilon)}
\frac{\Gamma(1-k-\epsilon)}{\Gamma(-k-\epsilon)}\frac{1}{\Gamma(k+1)}\nonumber\\
&=&\frac{\Gamma(k+1/2)}{\Gamma(1/2)}\frac{1}{-\epsilon}
\frac{\Gamma(1+k+\epsilon)}{\Gamma(k+\epsilon)}\frac{\sin(-k\pi-\epsilon\pi)}{\sin(\pi-k\pi-\epsilon\pi)}\frac{1}{\Gamma(k+1)}\nonumber\\
&=&\frac{1}{\epsilon}\frac{\Gamma(k+1/2)}{\sqrt{\pi}\ \Gamma(k)}.
\end{eqnarray}
Therefore, this coefficient is divergent as $\epsilon\rightarrow0$
and cancel the divergence of (\ref{AsymptoticA}). This means the
phase space potential (\ref{correctedpowerlawU1}) is still valid
for half integers $n=k+1/2$ with $k\geq1$.

\

\subsection{Coulomb potential $n=1/2$ }

According to identity (\ref{Gauss's theorem}) we have
\begin{eqnarray}\label{}
&&_2F_1(\frac{1}{2},\frac{1}{2}-n;\frac{3}{2}-n;1)\nonumber\\
&=&\frac{\Gamma(\frac{3}{2}-n)\Gamma(\frac{1}{2})}{\Gamma(1-n)\Gamma(1)}\nonumber\\
&=&-\sqrt{\pi}\
\frac{\Gamma(n)}{\Gamma(n-\frac{1}{2})}\frac{\sin(n\pi)}{\cos(n\pi)}.
\end{eqnarray}
If we assume $n=1/2+\epsilon/2$ with $\epsilon\rightarrow0$, we
have
\begin{eqnarray}\label{}
&&_2F_1(\frac{1}{2},\frac{1}{2}-n;\frac{3}{2}-n;1)\nonumber\\
&=&
_2F_1(\frac{1}{2},-\frac{\epsilon}{2};1-\frac{\epsilon}{2};1)\nonumber\\
&=&\frac{\Gamma(1-\frac{\epsilon}{2})\Gamma(\frac{1}{2})}{\Gamma(\frac{1}{2}-\frac{\epsilon}{2})\Gamma(1)}\nonumber\\
&=&\sqrt{\pi}\
\frac{\Gamma(\frac{1}{2}+\frac{\epsilon}{2})}{\Gamma(\frac{\epsilon}{2})}\frac{\cos(\pi\epsilon/2)}{\sin(\pi\epsilon/2)}\nonumber\\
&\rightarrow&\frac{1}{\sqrt{\pi}}\frac{\Gamma(\frac{1}{2}+\frac{\epsilon}{2})}{\frac{\epsilon}{2}\Gamma(\frac{\epsilon}{2})}\nonumber\\
&=&\frac{1}{\sqrt{\pi}}\frac{\Gamma(\frac{\epsilon}{2})\Gamma(\frac{1}{2}+\frac{\epsilon}{2})}{\frac{\epsilon}{2}\Gamma(\frac{\epsilon}{2})^2}\nonumber\\
&=&\frac{2^{1-\epsilon}\Gamma(\epsilon)}{\frac{\epsilon}{2}\Gamma(\frac{\epsilon}{2})^2}\nonumber\\
&\approx&2^{1-\epsilon}\frac{\frac{1}{\epsilon}e^{-\gamma
\epsilon}}{\frac{\epsilon}{2}(\frac{\epsilon}{2})^{-2}e^{-\gamma
\epsilon}}\nonumber\\
&=&2^{-\epsilon}\nonumber\\
&\approx&1-\epsilon\ln2.
\end{eqnarray}
Using the above formula, we have $U(R_{ij})$ for the potential
$V(x_i-x_j)=\beta|x_i-x_j|^{-1-\epsilon}$,
\begin{eqnarray}\label{powerlawV3}
U(R_{ij})&=&\frac{2\beta}{\pi
R^{2n}_{ij}}\frac{t^{1-2n}}{1-2n}\ _2F_1(\frac{1}{2},\frac{1}{2}-n;\frac{3}{2}-n;t^2)\Big|_{\sin\tau_c}^1\nonumber\\
&=&\frac{2\beta}{\pi
R^{1+\epsilon}_{ij}}\frac{1}{-\epsilon}(1-\epsilon\ln2-\tau_c^{-\epsilon})\Big|_{\epsilon\rightarrow
0,\tau_c \ll 1}\nonumber\\
&=&\frac{2\beta}{\pi
R^{1+\epsilon}_{ij}}\frac{1}{-\epsilon}(1-\epsilon\ln2-1+\epsilon\tau_c^{-\epsilon}\ln\tau_c)\Big|_{\epsilon\rightarrow
0,\tau_c \ll 1}\nonumber\\
&=&\frac{2\beta}{\pi R_{ij}}\ln(2/\tau_c)\Big|_{\epsilon=0,\tau_c
\ll 1}.
\end{eqnarray}

\end{document}